\documentstyle{article}
\begin{document}

\baselineskip=23pt

\begin{center}
{\bf {\huge The hardness-duration correlation in the two classes of
gamma-ray bursts}}\\\vspace{4mm} {\bf Yi-Ping Qin$^{1,2,3,4}$, Guang-Zhong
Xie$^{1,2,4}$, Sui-Jian Xue$^{2,3}$, Xue-Tang Zheng$^5$, and Dong-Cheng Mei$%
^{1,2,4,6}$ }\\{\bf $^1$ Yunnan Observatory, Chinese Academy of Sciences,
Kunming, Yunnan 650011, P. R. China }\\{\bf $^2$ National Astronomical
Observatories, Chinese Academy of Sciences }\\{\bf $^3$ Chinese Academy of
Science-Peking University joint Beijing Astrophysical Center, Department of
Geophysics, Peking University, Beijing 100871, P. R. China}\\{\bf $^4$
Yunnan Astrophysics Center, Yunnan University, Kunming, Yunnan 650091, P. R.
China }\\{\bf $^5$ Department of Physics, Nanjing University of Science and
Technology, Nanjing, Jiangsu 210014, P. R. China }\\{\bf $^6$ Department of
Physics, Yunnan University, Kunming, Yunnan 650091, P. R. China }\\%
\vspace{6mm} {\Large Abstract}\\
\end{center}

The well-known hardness-duration correlation of gamma-ray bursts (GRBs) is
investigated with the data of the 4B catalog. We find that, while the
hardness ratio and the duration are obviously correlated for the entire set
of the 4B catalog, they are not at all correlated for the two subsets
divided at the duration of 2 seconds. However, for other subsets with
comparable sizes, the two quantities are significantly correlated. The
following conclusions are then reached: (1) the existence of two classes of
GRBs is confirmed; (2) the hardness ratio and the duration are not at all
correlated for any of the two classes; (3) different classes of GRBs have
different distributions of the hardness ratio and the duration and it is
this difference that causes the correlation between the two quantities for
the entire set of the bursts.

\begin{flushleft}
\end{flushleft}

\section{Introduction}

Since the discovery of the events of gamma-ray bursts (GRBs) about thirty
years ago (Klebesadel et al. 1973), many achievements have been obtained,
but the full comprehension of the objects seems still to be a longstanding
problem. Among the many efforts, investigating statistical properties of the
events is as necessary as poking into the details of the bursts. Since more
and more data of GRBs have been available (e.g., Fishman et al. 1994; Meegan
et al. 1994; Meegan et al. 1996; Meegan et al. 1998; Paciesas et al. 1999),
statistical results become more and more reliable. Possible correlations
among various parameters of GRBs were studied previously (e.g., Golenetskii
et al. 1983; Barat et al. 1984; Belli 1993). Investigations of the issues
were continued recently with more sizable sets of data (e.g., Mallozzi et
al. 1995; Dezalay et al. 1997; Belli 1999). With a large number of bursts
observed with BATSE, Fishman (1999) found that the hardness-duration
correlation, which had been described previously, was confirmed. In the
following, we will make a further investigation on this issue.

\section{The hardness-duration correlation of GRBs}

It is well-known that there are two classes of bursts with different
distributions of duration, divided at around 2 seconds (e.g., Dezalay et al.
1992; Kouveliotou et al. 1993; Fishman et al. 1994; Meegan et al. 1996;
Paciesas et al. 1999). We wonder if the hardness-duration correlation is
caused by different distributions of the hardness ratio and the duration for
the two classes. In other words, we want to know if the correlation is still
held for either of the two classes.

To investigate this issue, the burst data of the 4B catalog (Paciesas et al.
1997) are employed. We divide the bursts into two subsets with a division at
the duration of 2 seconds. That is, those bursts with $T_{90}<2s$ are
defined as the short duration bursts and those with $T_{90}\geq 2s$ are
defined as the long duration bursts. The hardness ratio of a burst is
defined as the fluence in channel 3 ($\sim 100$ to $\sim 300$keV) divided by
the fluence in channel 2 ($\sim 50$ to $\sim 100$keV). There are 1179 bursts
in the catalog with available values of $T_{90}$ and the fluences in both
channels 2 and 3. This set is called sample 1. Of the 1179 sources, 304
belong to the short duration burst class and 875 constitute the long
duration burst class. The two subsets are called samples 2 and 3,
respectively. The correlation between $\log HR$ and $\log T_{90}$ is
calculated for the three samples, where $HR$ denotes the hardness ratio
defined above. We find: (1) the correlation coefficient between the two
quantities for sample 1 is $r=-0.391$, where the size of the sample is $%
N=1179$; (2) for sample 2, $r=0.002$ where $N=304$; (3) for sample 3, $%
r=-0.050$ where $N=875$. This shows that, while the hardness ratio and the
duration are obviously correlated for the entire set of the 4B catalog, they
are not at all correlated for any of the two classes. The correlation shown
in sample 1 must be caused by the different distributions of the hardness
ratio and the duration of the two classes.

To get an intuitive view of this point, we make a plot of $\log HR-\log
T_{90}$ for the sources (shown in Figure 1). In the plot, all data points
are presented and the regression lines for the three samples are drew.
Presented in the plot are also two data points standing for the average
values of the two quantities for the two classes. A straight line connecting
these two data points is drew. We find in Figure 1 that, the regression line
of the entire set of the 4B catalog is very close to the straight line, but
obviously deviates from the two other regression lines, suggesting that the
correlation shown in sample 1 is indeed caused by the different
distributions of the two classes.

\section{Discussion and conclusions}

In last section, we investigate if the well-known hardness-duration
correlation is caused by different distributions of the hardness ratio and
the duration for the two classes of GRBs. To investigate this issue, we
employ the burst data of the 4B catalog (Paciesas et al. 1997) and divide
the bursts into two subsets with a division at the duration of 2 seconds. We
find that, while the hardness ratio and the duration are obviously
correlated for the entire set of the 4B catalog, they are not at all
correlated for any of the two classes. The correlation shown in sample 1
must be caused by the different distributions of the hardness ratio and the
duration of the two classes.

Before reaching a conclusion, we must make clear if any subsets of the
catalog would produce an incorrelation between the two quantities. Firstly,
we select a subset of the 4B catalog by constraining the duration in the
range $1\leq T_{90}<10$. This subset contains 217 sources (called sample 4).
For sample 4, the correlation coefficient between the two quantities is $%
r=-0.343$, where $N=217$. Secondly, we select another subset of the 4B
catalog by constraining the duration in the range $0.1\leq T_{90}<100$ (note
that the range of the duration for the entire set of the 4B catalog is $%
0.01\leq T_{90}<1000$). We then get a 1045 source sample (called sample 5).
The correlation coefficient between the two quantities for sample 5 is $%
r=-0.413$, where $N=1045$. The two quantities are significantly correlated
for these two subsets. One should notice that: the sizes of samples 4 and 5
are comparable with that of samples 2 and 3 but the two former contain the 2
second division point. This clearly indicates that only for those subsets
belonging to one of the two classes of GRBs there would show an
incorrelation between the two quantities; subsets containing big enough
numbers of sources of both classes would present an obvious correlation
between the two quantities. This analysis not only reinforces the above
conclusion, but, in turn, also confirms the existence of two classes of GRBs.

We then come to the conclusion that, there indeed exist two classes of GRBs;
the hardness ratio and the duration concerned are not at all correlated for
any of the two classes; different classes of GRBs have different
distributions of the hardness ratio and the duration and it is this
difference that causes the correlation between the two quantities for the
entire set of the bursts.

\vspace{30mm}

\begin{center}
{\bf {\Large Acknowledgments}}\\
\end{center}

This work was supported by the United Laboratory of Optical Astronomy, CAS,
the Natural Science Foundation of China, and the Natural Science Foundation
of Yunnan.

\newpage

\begin{center}
{\bf {\Large References}}\\
\end{center}

\begin{verse}
Barat, C., et al. 1984, ApJ, 285, 791 \\Belli, B. M. 1993, A\&A, 97, 63 \\%
Belli, B. M. 1999, A\&AS, 138, 415 \\Dezalay, J.-P., et al. 1992, Proc.
Huntsville Gamma-Ray Burst Workshop, ed. W. S. Paciesas \& G. J. Fishman
(New York: AIP), 304 \\Dezalay, J.-P., et al. 1997, ApJ, 490, L17 \\Fishman,
G. J. 1999, A\&AS, 138, 395 \\Fishman, G. J., et al. 1994, ApJS, 92, 229 \\%
Golenetskii, S. V., et al. 1983, Nature, 306, 451 \\Klebesadel, R., Strong,
I., Olson, R. 1973, ApJ, 182, L85 \\Kouveliotou, C., et al. 1993, ApJ, 413,
L101 \\Mallozzi, R. S., et al. 1995, ApJ, 454, 597 \\Meegan, C. A., et al.
1994, The Second BATSE Burst Catalog, available electronically from the
Compton Observatory Science Support Center \\Meegan, C. A., et al. 1996,
ApJS, 106, 65 \\Meegan, C. A., et al. 1998, in AIP Conf. Proc. 428,
Gamma-Ray Bursts: 4th Huntsville Symp., ed. C. A. Meegan, R. D. Preece, \&
T. M. Koshut (New York: AIP), 3 \\Paciesas, W. S., et al. 1997, The Fourth
BATSE Burst Catalog, available electronically at
http://cossc.gsfc.nasa.gov/cossc/batse/4Bcatalog \\Paciesas, W. S., et al.
1999, ApJS, 122, 465 \\
\end{verse}

\vspace{2mm}

\begin{center}
{\bf {\Large Figure Caption}}\\
\end{center}

\begin{verse}
{\bf Figure 1.} The plot of $\log HR-\log T_{90}$ for sample 1 (including
samples 2 and 3), where, $T_{90}$ is in units of s. In the plot, a plus
overlapping an open square represents a source of sample 2, while an open
circle stands for a source of sample 3. The solid line is the regression
line for sample 1, while the two dotted lines are the regression lines for
samples 2 and 3, respectively. Filled circles represent the two data points
standing for the average values of the two quantities for the two classes
respectively. The dash line is a straight line connecting these two data
points.
\end{verse}

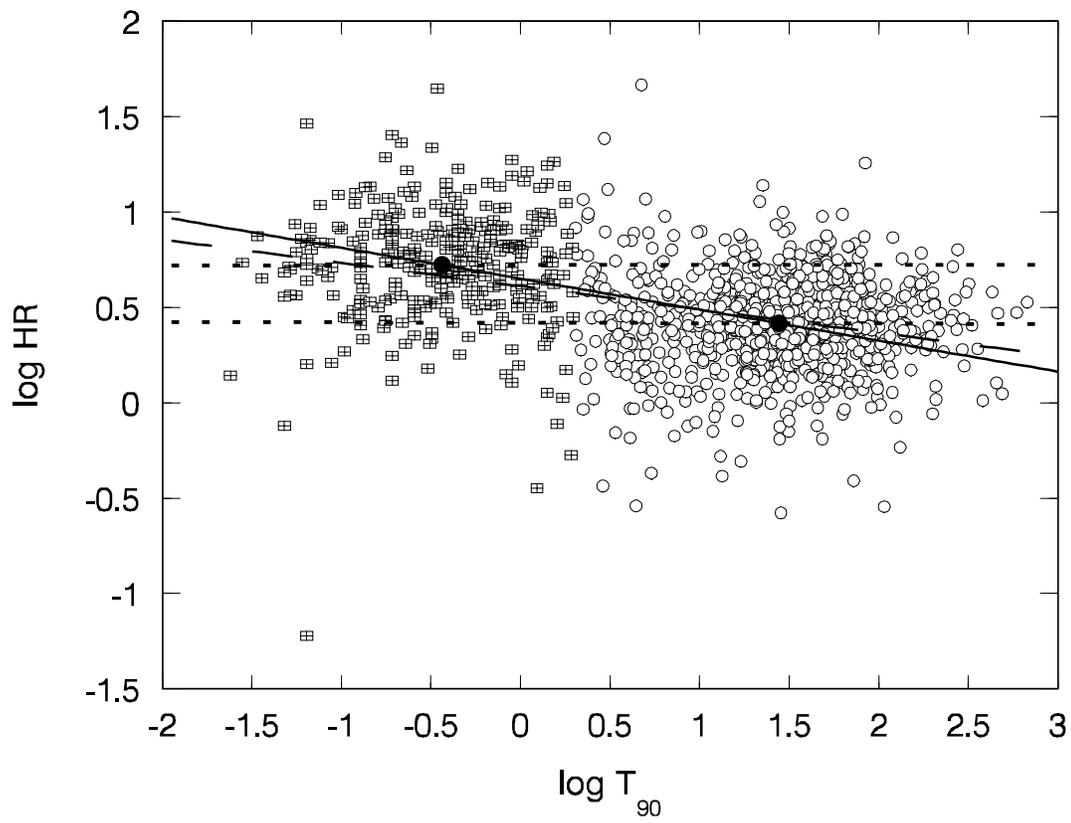
\end{document}